# Architecting Safer Autonomous Aviation Systems[†]

Jane Fenn[1], Mark Nicholson[2], Ganesh Pai[3], and Michael Wilkinson[1]

[1]BAE Systems, UK
{jane.fenn, michael.wilkinson1}@baesystems.com

[2]University of York, UK
mark.nicholson@york.ac.uk

[3]KBR / NASA Ames Research Center, USA
ganesh.pai@nasa.gov

**Abstract**

*The aviation literature gives relatively little guidance to practitioners about the specifics of architecting systems for safety, particularly the impact of architecture on allocating safety requirements, or the relative ease of system assurance resulting from system or subsystem level architectural choices. As an exemplar, this paper considers common* architectural patterns *used within traditional aviation systems and explores their safety and safety assurance implications when applied in the context of integrating artificial intelligence* (AI) *and machine learning* (ML) *based functionality. Considering safety as an architectural property, we discuss both the allocation of safety requirements and the architectural trade-offs involved early in the design lifecycle. This approach could be extended to other assured properties, similar to safety, such as security. We conclude with a discussion of the safety considerations that emerge in the context of candidate architectural patterns that have been proposed in the recent literature for enabling autonomy capabilities by integrating AI and ML. A recommendation is made for the generation of a property-driven architectural pattern catalogue.*

## 1 Introduction

Architecture and architecting have wide application in systems engineering, with architecture definition reportedly being one of the most often-used processes in model-based systems engineering (MBSE) (Cloutier and Bone 2015). Despite this, there is relatively little guidance available to practitioners on how to devise an architecture for a specific purpose, such as the incorporation of untrusted, but novel technologies, e.g., artificial intelligence (AI) and machine learning (ML), within a safety-critical system. Within the aviation sector, and beyond, there has been considerable interest in the application of AI/ML to achieve autonomous operation. A key obstacle is the nexus of issues surrounding safety, including the assurance of safe autonomous operation, which is particularly acute when enabling technologies such as AI/ML may be non-deterministic and/or unpredictable at worst, and complex and opaque, at best.

In aviation, safety is the 'state in which risk is acceptable'. Practitioners in the domain recognise that choices made by system designers and implementers will have significant impact on the safety of a system, both on the allocation of requirements across the elements of the system design and also on how assurance of the implemented system is achieved. An architecture is commonly understood as the *organisation* or *structure* of a system in terms of its elements and their relationships (ISO/IEC/IEEE 2011, ISO/IEC/IEEE 2015).

---

[†]This work was co-authored by an employee of KBR, Inc. under Contract No. 80ARC020D0010 with the National Aeronautics and Space Administration. The United States Government retains and the publisher, by accepting the article for publication, acknowledges that the United States Government retains a non-exclusive, paid-up, irrevocable, worldwide license to reproduce, prepare derivative works, distribute copies to the public, and perform publicly and display publicly, or allow others to do so, for United States Government purposes. All other rights are reserved by the copyright owner.

This paper studies architectural structures that combine untrusted elements with trusted elements in such a way that the overall system can be considered safe. In this context, *safety* is an abstract property, which needs to be interpreted precisely in the context of the architectural structure proposed for a system.

Our paper is motivated by considerations of whether the way we currently architect aviation systems could help ensure that systems using AI/ML components are safe by design. We address these questions by assessing a set of existing and newly proposed standard architectural forms, *architectural patterns*, when AI/ML components are employed as part of the architecture[1]. We also give a reminder of the principles, objectives, and practices for architecture development in aviation as much of this knowledge is not explicit in current de facto standards (SAE 1996b), (SAE and EUROCAE 2010); rather it is implicit knowledge amongst the authors of those standards and guidance documents, and the practitioners in the domain.

We hope this paper presents useful and interesting perspectives for many readers, but we anticipate the content being of particular interest to established safety engineering practitioners who are beginning to look at the issues with integrating AI/ML into their systems, and also to AI/ML development professionals who recognise that integrating their novel technology requires consideration of system safety. To appeal to as diverse an audience as possible, we avoid a presumption of a high level of prior knowledge and necessarily constrain our assessment to basic architectural patterns, deriving our observations from basic principles.

## 2 Background

### 2.1 Basic Concepts

Architecture concepts can be applied to any kind of system, at any level of system breakdown, and from any perspective (Wilkinson and Rabbets 2020). Typically, certain perspectives are used to draw out types of architectural structure in both problem and solutions spaces, ranging from operational, through logical/functional to system/physical. Safety concerns may need to be addressed within each of these types of architectural structure.

A common assumption is that architects work 'top down': i.e., operational need informs the logical or functional structure that, in turn, determines the system or physical structure. In practice, this is often not the case. In this paper, we assume a new solution-space technology choice (AI/ML) has been selected. We then consider a set of logical or functional architectural patterns that could be employed and how AI/ML-based functions could be linked 'up' to the operational structure and 'down' to the system/physical structure.

In the aviation domain, an aircraft is a *product system* (INCOSE 2022) that exists within a broader system of systems. Each aircraft comprises interacting systems, themselves composed of subsystems and/or *line replacement units* (LRUs), which may have connections to resources such as sensors, effectors, electrical power, and cooling air. Within an aircraft, some avionic systems may employ a prescribed generic logical and physical architecture with defined interfaces between elements, e.g., *integrated modular avionics* (IMA), *full airborne capability environment* (FACE), and *common avionics architecture system* (CAAS).

A logical or functional architecture might indicate how functions and behaviour are allocated within a system element (e.g., LRU), paying attention to issues such as independence requirements. A physical architecture might show which components may be used. Where a system element contains diverse technologies, such as software and hardware in programmable devices, there will be distinct but related software and hardware architectures which together provide the overall architecture for an element of the system.

An AI/ML technology constitutes a specific type of programmable element; one whose behaviour may in general be non-deterministic, or at best be less transparent than conventional software and complex hardware. The precise nature of the challenges to be addressed architecturally depend on a multitude of factors, including the amenability of the element to analysis, its variability depending on learning algorithms and data, and its behaviour under anomalous conditions.

---

[1] Though we refer to AI/ML broadly in this paper, we specifically consider neural networks (NNs), a particular form of ML. We acknowledge that the terms *autonomy*, *artificial intelligence*, and *machine learning* have distinct meanings and implications, although they are often used interchangeably.

## 2.2 Architecture Principles

A system architecture can be used to establish an understanding of how the system is organised, e.g., how its constituent elements are structured, their boundaries, and the boundary to the environment. Additionally, it can serve to reconcile and realise competing requirements into a feasible basis to guide system design, development, and evolution. From a safety standpoint, the system architecture supports the principles and design decisions by which the emergent behaviour of the system (and that of its constituent elements) can be constrained to an acceptably safe operational state space.

Producing an architecture that is fit for purpose is a creative activity. In practice, architects use their experience to *hypothesise and test*: that is, define candidate architectural structures, informed by known requirements and constraints, and then assess whether those structures exhibit desired properties.

However, over the years architectural principles have emerged; for example, eleven principles or techniques for *fail-safe design* have been introduced in (FAA 1988, EASA 2021a) for the safety characteristics of aircraft. These include strategies such as redundancy or backup systems, warnings and error-tolerance, and the corresponding guidance suggests that a combination of two or more of the principles may usually be needed to achieve a fail-safe design. Wider dependability (reliability, availability, etc.) aspects also need to be considered as part of the architecting process.

A typical system architecture exhibits the following characteristics:
- the system organisation, boundaries, interfaces, and behaviours are refined to an appropriately low level[2];
- all established requirements (i.e., functional, performance, integrity, reliability, safety, security, robustness, and derived requirements, as well as those pertaining to interfaces, integration, and those stemming from physical, environmental, technological, and implementation constraints) are appropriately allocated;
- the stated system and safety requirements are satisfied;
- the following principles are leveraged as appropriate:
    - independence (e.g., in the safety-related features)
    - diversity (functional, physical, design, data)
    - layering, i.e., defence in depth
    - avoidance, detection, and containment (e.g., through mechanisms such as redundancy, cross-checking, isolation, masking, etc.)

## 2.3 Developing an Architecture

There is ongoing work to provide guidance on safety assurance when integrating AI/ML into aviation systems (EASA 2021b, SAE and EUROCAE 2022) that outlines some candidate objectives, albeit from a process assurance standpoint. In (SASWG 2022), specific objectives for architecture have been explicitly put forth as applicable in the context of autonomy, although process guidance has (intentionally) not been given.

For this paper, the architectural challenge is to accommodate elements incorporating currently untrusted technology (AI/ML) within an overall architecture that possesses adequate safety, defined in some way that is meaningful to stakeholders. In general this involves hypothesising an architecture and testing it, usually by means of analysis or argumentation of an appropriate kind, to ensure that it is safe.

Typically, an analyst would assess the safety impact of a range of failure modes using methods such as *Functional Hazard Assessment* (FHA), as in ARP 4761 (SAE 1996b). For software, limited guidance is available on the nature of failure modes to be considered.

A number of enhanced approaches have been proposed that are more tailored for complex software-intensive systems, such as *Software Hazard Analysis and Resolution in Design* (SHARD) (Pumfrey 1999), which employ a set of *guidewords* to establish software contribution to system safety. One such set is:
- *Omission*: The service is never delivered, i.e., there is no communication.
- *Commission*: A service is delivered when not required, i.e., there is an unexpected communication.
- *Early*: The service (communication) occurs earlier than intended. This may be absolute (i.e., early compared to a real-time deadline) or relative (early with respect to other events or communications in the system).
- *Late*: The service (communication) occurs later than intended. As with early, this may be absolute or relative.
- *Value*: The information (data) delivered has the wrong value.

These are useful concepts to consider when the system context is understood. However, analysts may require additional guidance on the interpretation of those guidewords and, indeed, additional guidewords may be appropriate

---
[2] i.e., An *item* level, based on the terminology used in civil aviation.

when using AI/ML. We recommend this as future work, outside the scope of this paper.

Other techniques have also been proposed which take a broader perspective of the system in context of its operating environment (Leveson 2016). They similarly analyse deviations on expected performance at the system architecture level and assesses the potential safety impact.

Once safety requirements have been generated, defining an architecture usually involves function decomposition, trade-off analyses, consideration of design principles, as well as decomposition into lower levels of the system hierarchy. These activities should all be undertaken in the context of an assessment of the credibility of the design from a safety standpoint.

There are numerous potential approaches to aiding architectural selection such as the *Architecture Trade-off Analysis Method* (ATAM) (Kazman et al. 2000), and *Trade Trees* (NASA 2017). In general, these methodologies start by defining the scope or trade space, e.g., by defining the drivers for an architectural choice, such as tolerance to change, and various (prioritised) scenarios or configurations that characterise those drivers.

Then, the architectural options are identified together with a definition and prioritisation/weighting of the attributes of interest (e.g., reliability, safety, and security). Each option is analysed against each identified attribute and for each scenario, towards identifying (and selecting) optimised solutions.

Trade-offs are important drivers. For example, safety attributes may relate to not only the perceived ease in implementing safety requirements derived as a result of the architecture, but also the ease of generating the assurance evidence to demonstrate that the requirements have been met. Inspiration for hypothesising an architecture for a system may be drawn from previously successful architectures, captured as *architectural patterns*.

## 3 Architectural Patterns

The use of *patterns* has been recognised as a way of capturing good practice for many decades, initially in architecture, and subsequently, for system and software design, including successful safety approaches.

As software became routinely introduced into the development of systems whose behaviour could result in harm to humans, patterns of system-level safety protection architectures previously developed for hardware were updated. As software was introduced into aircraft systems, what were then new, and are now established, architectural patterns emerged. The initial version of the aerospace recommended practice for development of civil aircraft, ARP 4754 (SAE 1996a), lists the following patterns considered relevant to aerospace applications: (i) *partitioned design*; (ii) *dissimilar, independent designs implementing an aircraft-level function*; (iii) *dissimilar designs implementing an aircraft-level function*; (iv) *active-monitor parallel design*; and (v) *backup parallel design*. In much the same way, new architectures and architectural patterns are being proposed to integrate AI in general and ML in particular into safety-critical systems, especially in aviation.

What are the practical implications for selecting a system architecture with the move to integrating ML/AI? We revisit both the established and new architectural patterns, and investigate their suitability for use. We start by reviewing the simplest architecture: *single channel design* (Section 3.1.1). Then we examine the potential impacts when using two of the established patterns, active-monitor parallel design (Section 3.1.2) and backup parallel design (Section 3.1.3), with an AI/ML element, considering the contribution to system safety. We also examine how combining these established patterns (Section 3.1.4) relates to the newer proposed patterns (Section 3.2) and the considerations that emerge when including AI/ML based functions.

For what follows, we use *complex function* to mean an AI/ML-based complex function and will qualify the term when it is not clear from context.

### 3.1 Generic Architectural Patterns in Aircraft Systems

#### 3.1.1 Single Channel Design

The single channel design (Fig. 1) is the simplest architecture possible for a system, where the complex element (originally software or complex hardware) inherits the totality of the safety requirements allocated to the function that it implements, in terms of failure probabilities (or failure rates) and assurance[3] requirements.

Failure rate requirements are expressed per average operational hour, while assurance requirements at a system level are determined (in civil aviation) from *Development Assurance Levels* (DALs)[4], i.e., levels A–E, with level A representing the highest criticality level and mapping to the most stringent requirements on process rigour.

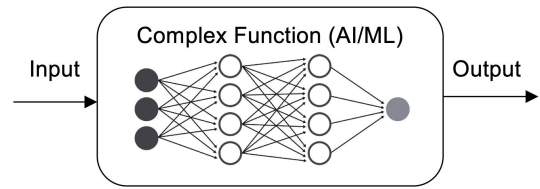

**Fig. 1.** Single channel design

**Discussion.** When the complex function in this pattern is implemented using AI/ML in general, and deep-learning in particular, it captures a so-called *end-to-end learning* architecture (Bojarski et al. 2016).

At present, there are neither broadly accepted techniques for determining failure rates for ML component failure modes that potentially contribute to system safety[5], nor is there an agreed approach to demonstrating achieved assurance. The probabilistic requirements for safety-critical systems are stringent, and it is not clear that current ML techniques can meet these requirements. For applications with minimal levels of safety requirements placed on the system, perhaps through their limited scope of use, the single channel architecture may present an acceptable residual risk. It is unlikely to be acceptable for moderate and higher levels of safety requirements allocated to a function.

### 3.1.2 Active-Monitor Parallel Design

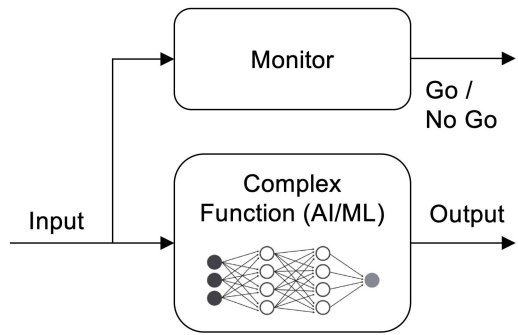

**Fig. 2.** Active-Monitor Parallel Design

*Active-monitor parallel design* (Fig. 2) is shown in the simplest, most generic form of architectural pattern. It requires a degree of interpretation for practical implementation: the monitor might be used either to disconnect the output from the complex function, or otherwise indicate the status of the output as *invalid* to downstream elements that consume the output. This pattern is predominantly about handling *erroneous function* or *malfunction* of the complex element, e.g., failure modes of *value* or *timing* that may have been derived, for example, using SHARD guidewords (see Section 2.3).

---

[3] *Assurance* in this context means the approach used to manage systematic errors, which, in civil aviation, is usually associated with process rigour. In general, the term may be used slightly differently in other domains and in the associated standards/regulations.

[4] Functions are assigned so-called *function* DALs (FDALs). Upon decomposition and allocation to items, *item* DALs (IDALs) determine item-level assurance requirements. IDALs for software items are equivalent to (software) *design assurance levels* (DALs), the terminology used in the guidance documents for software assurance in aircraft system certification (RTCA and EUROCAE 2012).

[5] (Cluzeau et al. 2020) have claimed being able to determine ML failure rates in a safety-critical aviation application, although they have withheld material explanation of the underlying methodology from wider public scrutiny.

**Discussion.** In the simplest form, the monitor understands the transfer function between input and output of the complex element and uses the inputs to ascertain whether the output would be valid. Consider examples such as input range checking. In terms of allocating safety requirements, this pattern implicitly assumes that *loss of function* is of less concern. In terms of protection against erroneous function, subject to meeting availability requirements, a typical usage (SAE 2010) could be to allocate the highest assurance requirements (i.e., DAL A) to the monitor, to be implemented with simple, more readily verifiable technologies, while the more complex function then has to meet less onerous assurance requirements, e.g., those mapped to DAL C.

When an (AI/ML-based) complex function is used in this pattern, additional considerations are necessary around the monitoring function. Simplistic range checking may be required but may not be sufficient to determine whether the output of the complex function is safe in its system context. Sometimes AI/ML performance may be inadequate from a safety perspective within some parts of the operating domain. For example, the monitor needs to flag the outputs invalid at those parts. In this case, additional monitoring could be introduced at the risk of increased complexity, which may also make verification of the monitor more challenging. We will revisit these issues when we discuss the *runtime assurance* pattern (Section 3.2.1).

### 3.1.3 Backup Parallel Design

*Backup parallel design* (Fig. 3) is another simple form of architectural pattern that is helpful to ensure availability and to protect against a *loss of function* failure condition of the complex function.

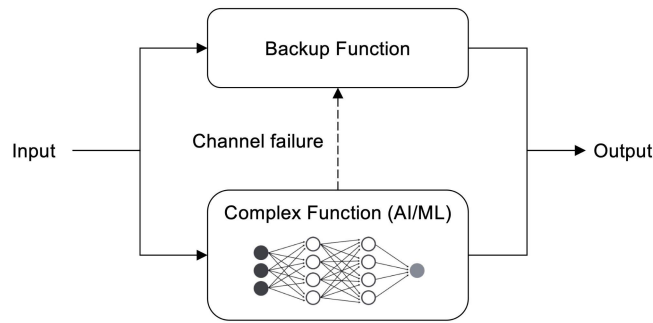

**Fig. 3.** Backup Parallel Design

Here, the obligation would be on the complex element itself to detect it is generating no output, or erroneous outputs, and self-suspend its function, as well as alert the backup function to take over. This suggests that it is possible to identify when the complex function has failed with a high degree of accuracy and certainty.

For safety requirements allocation, ARP4754A (SAE 2010) allows for DAL A requirements to be allocated such that, if sufficient independence could be shown, the primary (complex function) portion is implemented at DAL A and the backup portion at DAL C.

**Discussion.** When introducing AI/ML in the complex function, this allocation could be reversed so that the primary is allocated DAL C, with the backup allocated DAL A. Here, additional considerations emerge around both the self-test/self-diagnosis capability, and the balance between primary and backup functions. Self-diagnosis, and particularly detection of erroneous behaviour can become more complex than for traditional software. Hence, the patterns in the rest of this paper consider *channel error* rather than *channel failure*. Although the complex function may include non-ML elements such as pre- or post-processing within its boundary, to our knowledge limited evidence is available to conclude that those elements include self-diagnosis capabilities, for example, when AI/ML is used in perception pipelines.

The nature of the backup system requires specific considerations when using AI/ML. The pattern assumes an acceptable level of availability of the complex function, and one choice for the backup is functional equivalence to the complex element. Context-dependent decisions are necessary on whether the 'quality' of the backup function needs to be equivalent, or whether a 'limp home' *gracefully degraded* capability may be sufficient instead. The acceptability of such an approach will also be dependent on the anticipated balance of when the backup function will be the operational portion, based on how frequently the complex function reports as failed/invalid.

### 3.1.4 Combined Architectural Patterns

Industrial practice rapidly found it to be impractical to use simple patterns on their own, and currently it is more common to use a combination of patterns. We note that loss of function is typically more readily detected than erroneous function in conventional complex systems, and anticipate the same to be true for systems which will use AI/ML.

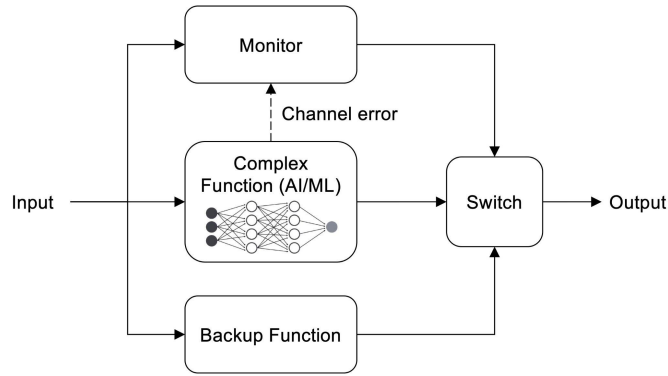

**Fig. 4.** A candidate combination pattern of active-monitor parallel design with backup parallel design

Based on the earlier discussions around the limited ability of AI/ML to detect its own failures, we now examine whether a combination of the active-monitor parallel design and the backup parallel design patterns could be useful to address the shortfalls of the individual patterns.

Several design choices are necessary when combining patterns to further address their individual shortcomings, so we illustrate an indicative implementation (Fig. 4) and consider a number of variants.

**Discussion.** For the pattern shown in Fig. 4, if the monitor determines that the complex function will not operate correctly, it switches to a conventionally assured backup function, assuming that an appropriate monitor can be constructed (see Section 3.2.1 for a more detailed discussion on this issue). As before, current practice would be to assign the overall pattern with safety targets in terms of failure rates and DALs. We assume that availability has been sufficiently addressed, and that erroneous function should be our main consideration.

When the pattern is used for conventional complex functions, the monitor and backup function inherit the same assurance requirements as the function allocated to the overall pattern. However, as previously mentioned, there are no broadly accepted techniques to determine the reliability of AI/ML-based complex functions, hence they need to be allocated lower target failure rates. In current practice, if the monitor and backup were allocated DAL A requirements for example, DAL C could be allocated to the (traditional software) complex function.

This allocation is not so straightforward for AI/ML-based complex functions. The combined architectural pattern of Fig. 4 requires that: (i) either the monitor understands the transfer function between inputs and outputs of the AI/ML-based complex function sufficiently to allow action to be taken when inputs are, for example, out of bounds; or (ii) the complex function can self-report erroneous function or loss of function. Where the monitor needs to switch to the backup, for instance in those regions of the *operational domain*[6] where the complex function performance has been determined to be inadequate, safety must be determined within the context of use.

If the complex function was introduced to improve capability, it is reasonable to infer that the backup function may not exactly replicate the ML-based function, and that the former's performance is an approximation of what the latter's output will be, either in the value domain, the timing domain, or both. In such cases, the switching behaviour at the appropriate portions of the input (i.e., the relevant regions of the operating domain), may itself be a safety property.

Moreover, it could be one that is defined during the development of the AI/ML-based complex function, with a lower level of rigour than that which would be necessary to support the overall allocated DAL for the monitor.

There are ramifications here for the cost of implementing this pattern. Current approaches assume that monitor development and assurance is feasible at a significantly lower cost and effort, relative to the complex function. The discussion above suggests that this is likely to be non-trivial. Also, as with the backup parallel design, the balance between the primary and backup functions needs to be considered for the pattern in Fig. 4. Next, we discuss whether the concerns, above, regarding the monitor could be addressed by using it differently.

---

[6]We use 'operational domain' to mean the *Operating Domain Model*, or *Operational Design Domain* (*ODD*), which are the (descriptions of the) operational domains in which the AI/ML-based complex function is (to be) designed to properly function.

**Variants of Pattern Combinations.** Using the monitor to observe the output of ML, rather than the inputs to the ML (Fig. 5) is one possible variation of the combined architectural pattern of Fig. 4.

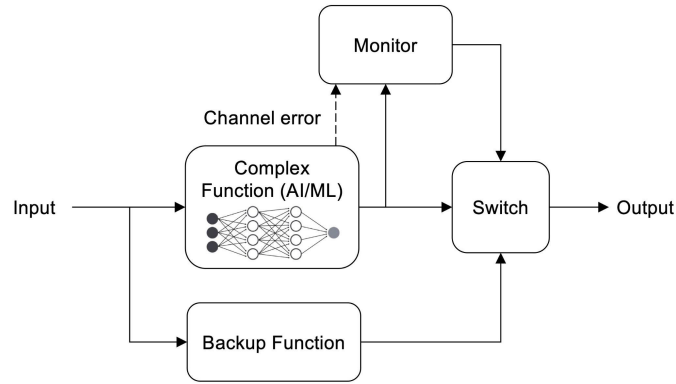

**Fig. 5.** Variant of the combined pattern of Fig. 4, with complex function *output* monitoring

In this configuration the monitor now judges whether the complex function output is valid against some defined criteria, relieving it of the requirement to know the transfer function implemented using AI/ML. However, the complexity is in defining the criteria for what constitutes *safe outputs*, particularly when the monitor has no understanding of the inputs that the complex function used to derive its outputs. Also, if the complex function is expected to process the inputs more rapidly than conventional means, for instance due to its optimisations and typical use of higher power and specialised computing hardware, there is a potential for a lack of synchronisation between the complex function output and what the monitor expects.

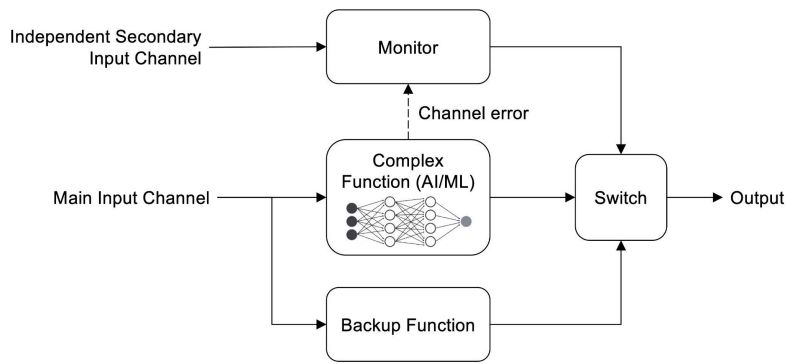

**Fig. 6.** Variant of the combined pattern of Fig. 4, with *independent channel input* monitoring

Another variation is to monitor the complex function inputs in addition to complex function outputs. In fact, this variant is the general configuration of the runtime assurance pattern discussed in more detail later (Section 3.2.1).

A third possible variant (Fig. 6) is to provide alternative inputs to the monitor. There is an independent sensing channel to observe environment conditions that are known a priori to degrade ML performance below acceptable levels. For instance, when using a complex function that was trained in high visibility conditions for vision-based perception in low light or low visibility conditions. In other words, the monitor checks for conformance to the operational domain defined for the complex function. This may provide additional confidence, providing the factors that cause poor ML performance are satisfactorily understood, with the rigour necessary to support the assurance requirements allocated to the monitor.

*Model-centred assurance* (Jha et al. 2020), a new alternative architecture proposed to enable autonomy, critiques this variant of the pattern. It asserts that 'perception functions of the primary system will surely be better resourced and more capable than those of the monitors' and concludes that the monitors should rather use the same input channels to construct a model of the environment.

The model-centred assurance architecture does not conform to the patterns considered in this paper. As such, we do not discuss it further here, leaving its assessment from an aviation safety standpoint to future work. Instead, we look at *other* architectural patterns from the literature that have been recently proposed for integrating AI/ML-based functionality into high criticality applications to enable autonomy.

## 3.2 Architectural Patterns for AI and ML

We consider the following four patterns in this section: *runtime assurance* (RTA), *value overriding*, *function modification*, and *input partitioning and selection.* Each pattern is briefly described, and their suitability for use in aviation is discussed, primarily from a safety standpoint.

### 3.2.1 Runtime Assurance

*Runtime assurance* (RTA) or its variations (Schierman et al. 2020, ASTM 2021) realised in a *system-level simplex* structure (Bak et al. 2009) is an architectural pattern for safety assurance of systems containing complex functions that cannot be approved to the requisite assurance level, for example as part of an aircraft system certification process that relies upon traditional development assurance.

The RTA pattern (Fig. 7) involves assured *runtime (safety) monitors*, receiving trusted inputs, that observe a complex, less-trusted or untrusted function. Upon detecting conditions that can violate safety, e.g., invalid inputs, deviant function outputs, or errant execution traces, the monitors trigger switching to one or more assured alternative functions to maintain a safe system state. To use the pattern, the RTA *wrapper* constituting the monitor(s), switch, and alternative function(s) must be assured to a higher level than the complex function.

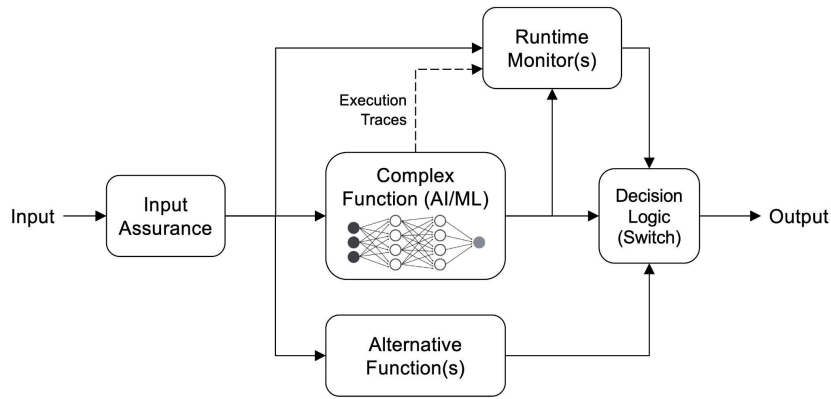

**Fig. 7.** Runtime assurance (RTA) pattern

There is compelling evidence that RTA works well in an aviation context (Burns et al. 2011). However, there are numerous considerations when using RTA for AI/ML-based functions, of which here we elaborate four.

**Choice of the Complex Function Boundary.** The AI/ML-based function boundary can have significant implications both on the assurance of the RTA scheme itself, and that of the integrated function (i.e., the AI/ML-based complex function secured within the RTA wrapper).

One possible boundary for the complex function includes pre-/post-processing computation besides the ML model (Fig. 8). Pre-/post-processing using conventional (software) development techniques is typical in AI/ML development and may involve operations that are comparable to those that occur in the monitor, e.g., input/output range checking, or handling null values. Thus, it may be tempting to include the monitor within the function boundary (Fig. 8) and provide it with pre-processed inputs rather than the *raw* input. However doing so poses conflicts for allocating assurance requirements. Either the complex function and the monitor require the same level of assurance, which would violate the safety intent of RTA, where the monitor has a higher level of assurance than the function being monitored;

or we must raise the level of assurance required for the complex function to that of the monitor, inducing a greater cost for generating the assurance evidence required.

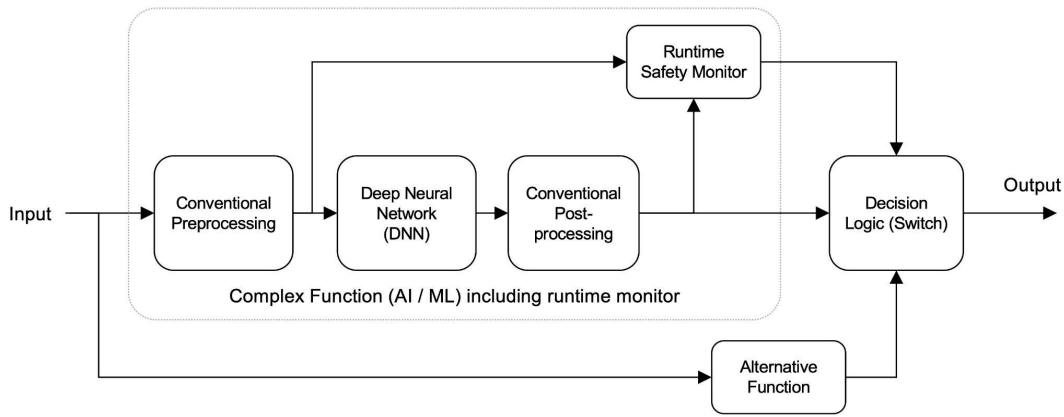

**Fig. 8.** RTA pattern including pre-/post-processing and the runtime monitor within the complex function boundary

Another possibility is to place only the ML model within the complex function boundary. In this case, either or both of pre- and post-processing will need the same level of assurance as the monitor.

Similarly, operations performed on the inputs necessary for high assurance could either be distributed across the monitors, complex functions, and the alternative functions(s), or be split off as a separate *signal conditioning*, or *input assurance* function (Fig. 7). In this case, other routine pre-processing, such as normalisation of data, would occur under lower assurance requirements within the complex function boundary.

**Monitor Function Considerations.** RTA requires a specification for the monitor that can be correctly implemented. This may be challenging in the context of AI/ML-based complex functions:
- Many monitors for detecting inputs that are not from the training distribution, i.e., *out-of-distribution* (OOD) inputs are themselves machine learnt. Similarly monitors for detecting out-of-operating domain inputs require assuming that the operating domain can be completely and comprehensively defined, which can be problematic for a perception function.
- Monitors to detect in-distribution inputs that could defeat the expected generalisation behaviour are difficult to specify and build because those inputs represent *surprises* that were previously unknown. Such inputs could be adversarial, or due to epistemic uncertainty about the operating domain.
- Depending on what the monitor is observing—e.g., raw inputs versus pre-processed inputs—value, timing, and synchronisation issues could potentially emerge that defeat the monitor, particularly while checking input/output validity.

**Assurance Requirements Allocation.** Where the RTA *wrapper* must itself be assured to the highest criticality level, i.e., DAL A, current development assurance processes require that:
1. the AI/ML function be assured to no lower than, in this case, DAL C. However, note that this pattern could be instantiated at different levels of abstraction, but assurance credit can only be taken *once* in the design. That is, the reduction in assurance to DAL C is therefore only permissible where no credit has been taken at a higher level of abstraction, e.g., at the system level;
2. all failure conditions of the AI/ML-based function need to be correctly detected under all operating conditions;
3. assurance is needed that no AI/ML-based failure condition can cause monitor failure.

Although condition three may be achievable in principle through independence, partitioning, or isolation strategies it is unclear whether or not it is feasible to satisfy the former two requirements for AI/ML-based functions.

**Configurations of RTA and Complex Functions.** The RTA pattern (Fig. 7) belies complicated configurations that involve multiple monitors and alternative functions including hierarchies of the same. In the general case, the decision logic involves decision-making under uncertainty which may, itself, be a complex (possibly AI/ML-based) function, thus making assurance of the RTA scheme at least as difficult as the assurance of the complex function. If the decision logic can be precisely specified, the safety assurance situation is more promising: it may be feasible to formally verify an AI/ML-based implementation of the same (Katz et al. 2017).

Apart from the monitor and backup functions, the complex function may itself be achieved using multiple simpler models rather than a monolithic model, as is common in *ensemble learning*. This might offer additional options for monitoring (Fig. 9) The individual models of the complex function may be simple and therefore monitoring those may also be simple. More analysis is needed however, to determine whether this simplifies or complicates monitoring overall, particularly due to issues of monitor consistency, timing, throughput, and synchronisation, for example.

As mentioned earlier (Section 3.1.4) the RTA pattern is, in principle, one variant of the combination of the active-monitor parallel design and backup parallel design patterns. Thus, many of the considerations applying to those simpler patterns also emerge here.

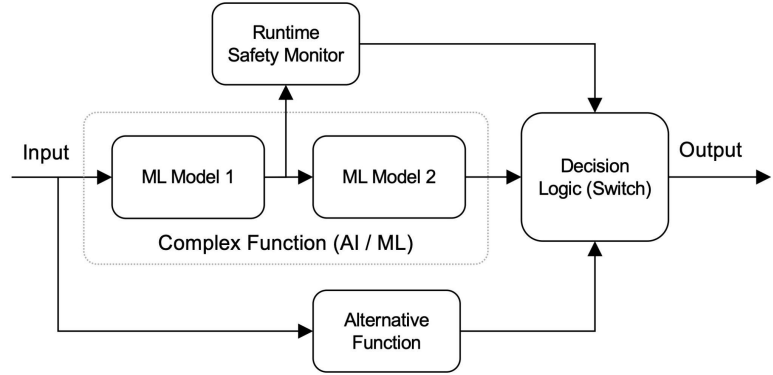

**Fig. 9.** RTA pattern including multiple ML models within the complex function

For instance, how often the alternative function is required to be operational, versus how frequently complex function failure is accepted; and the safety impact of the switching behaviour with respect to portions of the input domain where the complex function underperforms, and the corresponding timing requirements.

### 3.2.2 Value Overriding

In (Groß et al. 2022), four patterns have been proposed in the context of self-driving road vehicles for 'handling runtime uncertainty in perception', namely *uncertainty supervisor*, *safety margin selector*, *adaptive uncertainty supervisor*, and *adaptive safety margin selector*. Each of the proposed patterns may be applied to processing sensed data that have an associated uncertainty, e.g., an AI/ML model-based estimate of a quantity such as the coefficient of friction between vehicle tyres and the road surface, before they are presented as input to any subsequent computation.

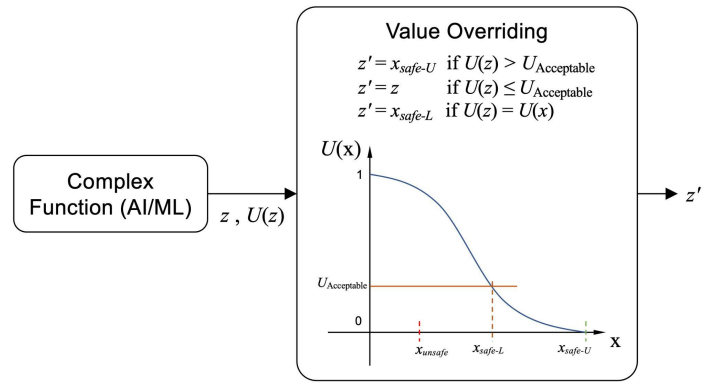

**Fig. 10.** Value overriding pattern (Non-adaptive)

The pattern to be used depends on how the data and its associated uncertainty are presented. The uncertainty supervisor pattern applies to point values, where a worst-case value replaces the input received when the uncertainty for that input exceeds a predetermined acceptable threshold. The safety margin selector pattern may be applied for input presented as a distribution, where the value replacing the input is, again, chosen using a predetermined threshold for acceptable uncertainty.

Effectively, these two patterns are variations of a common pattern of *value overriding* (Fig. 10) where the given data are overridden with *safe* values when their associated uncertainty exceeds some acceptable threshold. The *adaptive* variations of value overriding (i.e., adaptive uncertainty supervisor, and adaptive safety margin selector) vary the threshold for acceptable uncertainty itself, based on an additional input (not shown in Fig. 10) representing the (risk of the) operating situation.

**Discussion.** Applying the value overriding pattern induces additional requirements, not readily evident in (Groß et al. 2022):

- *worst-case safe values exist that can be independently established*: this requirement constrains the scope of the perception problem for which the AI/ML function is being used, and thus the kind of responses to which the pattern may be applied. In principle, this requirement may be achievable for quantities for which ground-truth worst-case reference values can be separately measured, determined through mathematical modelling, or using simulation.
- *reference uncertainty distributions used to select safe replacement values are valid and accurate*: as with the previous requirement, this requirement also could be met, in principle, through measurement, modelling, and simulation provided that there exists a ground-truth basis to the responses for which an uncertainty distribution is being determined.
- *uncertainty estimates produced by the AI/ML function can be trusted*: satisfying this constraint will be problematic when incorrect responses are produced with high confidence (equivalently, low uncertainty), e.g., in the presence of adversarial inputs to ML models, or unexpected inputs drawn from distributions differing from the training distribution.
- *operating situations are correctly determined*: although it may be possible in specific cases to use existing sensors and techniques for establishing the operating situation, in general this is itself the perception problem. Effectively, a circularity of requirements emerges where assured perception is needed to provide assurance of quantities themselves determined from perception.

The preceding requirements are not a comprehensive set and more analysis is needed to determine the additional requirements needed to usefully apply the value overriding pattern. For example, in conjunction with other architectural patterns, for different system configurations, and for different function criticalities.

By design, the value overriding pattern can modify the allowable safety margin in the responses of AI/ML-based functions. The justification for this choice is that (Groß et al. 2022):

> ...runtime estimation and handling of uncertainties is necessary to overcome worst-case approximations that would lead to unacceptable utility/performance, especially if the situation context indicates a low risk situation.

From an applicability in aviation standpoint, the principle underlying this pattern—that greater uncertainty in the responses from an AI/ML function is tolerable for performance gains in low(er) risk operating situations—violates the intent of the fail-safe design concept required by aircraft airworthiness regulations (FAA 1988, EASA 2021a). More specifically it violates *principle 10: margins or factors of safety to allow for any undefined or unforeseeable adverse conditions*. The rationale is that worst-case estimates for a quantity are rather a worst-case *lower bound*. Thus, reducing that safety margin introduces the additional burden of demonstrating that operating at any margin down to the alternative worst-case lower bound does not exhibit any unintended behaviour with unacceptable safety impact, under all foreseeable operating conditions. Furthermore, in aviation systems of systems (e.g., air traffic management (ATM), air navigation services (ANS), airport operations) safety margin reductions in favour of performance improvement can accumulate over time leading to *practical drift*, i.e., where safety performance is presumed safe but is in fact in an unforeseen, and appreciably higher-risk region than the approved baseline (ICAO 2018). As such, applying this pattern in its current form, especially for functions assigned higher DALs is unlikely to be acceptable.

### 3.2.3 Function Modification

Deep neural networks (DNNs) used for object detection as part of a vision-based perception system typically process sequences of input images and produce *bounding boxes* that spatially localise and highlight the detected objects of interest on each image. In use cases such as collision avoidance, the safety contribution of object detection to system hazards can be characterised by *false negative detections* (i.e., an object posing a collision hazard exists in the image, but the DNN does not recognise it) and *inaccurate localisations* (i.e., an object posing a collision hazard that exists in the image is correctly recognised, but the bounding box produced either partially covers it, or does not cover it).

For positive detections, *Intersection over Union* (IoU) is a frequently used metric of bounding box estimation performance, measuring the extent of overlap of the ground-truth and the predicted bounding boxes to quantify localisation accuracy. Thus, evidence from development that IoU is perfect (or nearly perfect) for all test input images not seen in training is desirable for safety assurance that the predicted bounding boxes will be accurate in deployment.

*Safety post-processing* (Schuster et al. 2022) has been put forth as a solution for this purpose in the automotive systems domain, for collision avoidance. Specifically, after conventional post-processing of the 2D, rectangular, and axis-aligned bounding boxes that the DNN produces on test images, safety post-processing scales them by an enlargement factor proportional to the IoU that was established during training. The possible range of values of the enlargement factor are mathematically proved to be the smallest required to guarantee that the enlarged bounding boxes will always contain the ground-truth object for detections on all input images, for all values of IoU obtained in training.

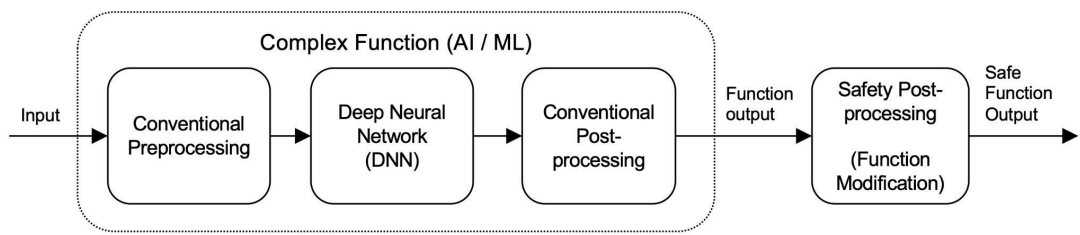

**Fig. 11.** Safety post-processing architecture, adapted from (Schuster et al. 2022), as an instance of the function modification architectural pattern

In general, this architecture could be seen as an instance of a pattern of (assured) *function modification* (Fig. 11) where the modified function outputs are (to be) guaranteed to meet the relevant safety constraints on the AI/ML function.

**Discussion.** The IoU metric admits values in the interval [0, 1], whereas the theorem that relates the enlargement factor to IoU (guaranteeing that ground-truth objects are always included in the scaled bounding boxes after safety post-processing) applies to the half-open interval (0, 1] (Schuster et al. 2022).
Since IoU = 0 for false positive detections, and neither true nor false negative detections produce bounding boxes, safety post-processing can only apply to the bounding boxes of true positive detections, and other mechanisms are required to minimise and mitigate false detections.[7] The latter may occur due to inadequate generalisation of learnt behaviour to nominal, out-of-sample, in-distribution inputs[8] (including edge and corner cases), insufficient robustness to adversarial, out-of-sample, in-distribution inputs, and/or OOD inputs that may be benign or adversarial.

In general, assurance that the learnt behaviour of the AI/ML function is robust and as expected over the required domain of in-distribution inputs is necessary for function modification to be usefully applicable. Additionally, applying the function modification pattern for perception, as shown in Fig. 11, in a single channel configuration (see Section 3.1.1) may not be sufficient for safety assurance. It requires combination with other architectural patterns. For example:
- a self-checking pair that includes diversity in the sensing and object detection functions, possibly in separate channels, to recognize false detections from DNN-based perception, with voting and/or sensor fusion to resolve disagreement in object detections;
- the active-monitor parallel design pattern that includes monitoring to detect out-of-distribution (OOD) inputs, coupled with OOD input handling.

---

[7] Here, although false positive detections do not have an immediate worst-case safety impact (since there are no obstacles posing a collision hazard), non-detection of a false positive either may be a precursor to a hazard—e.g., taking a corrective action when not required may itself be a hazard—or may lead to an effect of lower (but not insignificant) safety criticality, e.g., increased pilot or operator workload in an aviation context.

[8] Operational input data that are from the same distribution as the training and testing data (i.e., in-distribution), which were not sampled for inclusion into the training and testing required during the learning process (i.e., out-of-sample).

### 3.2.4 Input Partitioning and Selection

The *input partitioning and selection* architectural pattern (Fig. 12) contains two or more channels, each performing the same function, but on different partitions of the input domain of the function. One or more of the channels may include AI/ML, but at least one channel does not.

Conceptually, this pattern may be understood as a combination of a *demultiplexer* of the function input to different channels, and a *multiplexer* of different channels to the function output. The channel to be selected relies on monitoring of predefined conditions (e.g., whether the inputs are within certain bounds) and OOD inputs, or checking of predefined properties (e.g., whether the expected output is produced for a given input).

The 'hybrid' architecture proposed in (Damour et al. 2021) is an instance of this pattern containing 2 channels. The primary channel comprises a DNN-based implementation of the next-generation airborne collision avoidance system for unmanned aircraft (ACAS-Xu), which produces collision avoidance advisories for the given inputs. In fact, *multiple* DNN models are used to implement this complex function (an option previously considered in Section 3.2.1). The second channel is a lookup table (LUT)-based 'safety net' representing an alternative implementation of ACAS-Xu. This LUT-based channel is meant to operate on that portion of the input space for ACAS-Xu, where the DNN-based implementations are known to perform poorly, as established during their machine learning development lifecycle.

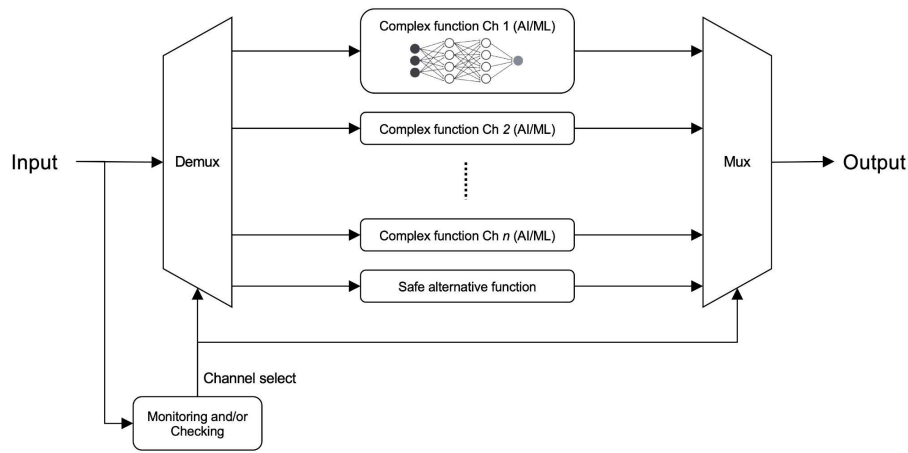

**Fig. 12.** Input partitioning and selection architectural pattern

**Discussion.** In its simplest incarnation, all channels are required to cover the complete input space for the function allocated to the pattern, and each channel operates only on a specific portion of that input space. Thus, at any given time, there is exactly one channel that can produce the required output for a given input, when that channel is operational. Loss of any channel thus leads to a loss of the function on the corresponding portion of the input.

For high criticality functions, high assurance is needed that: (i) the monitor/checking function invokes the safe alternative/backup function *only* in those portions of the operating domain where the primary channels and the backup are known to diverge; (ii) for the remaining portion of the operating domain, inputs (outputs) are correctly routed to (from) the appropriate channel; (iii) the safe backup is correct against a validated specification of the function allocated to the pattern; and (iv) the primary channel implemented using AI/ML is correct against the validated specification of the function for those portions of the operating domain where its responses are consistent with the safe backup.

There are numerous implications of the above assurance requirements. The first two conditions relate to correctness of the monitor/checking function. They additionally require the safe backup and the primary channel to be consistent in their outputs for some common portion of the operating domain, and that consistency be shown with high assurance. In the hybrid architecture (Damour et al. 2021), formal verification is used to establish this consistency property, although in general (e.g., for a perception function), this may be challenging to achieve or demonstrate.

The implication of condition three, above, is that the safe backup is itself a complex function albeit not one implemented using AI/ML, but using a conventional approach instead. In other words, it could in principle serve as a primary channel over the entire operating domain of the intended function. This necessarily requires complete functional equivalence of the AI/ML-based complex function and the safe backup, even if each is only being used on specific portions of the operating domain. Again, in general, this may be challenging to achieve, and might raise the (legitimate) question of needing to use an AI/ML implementation in the first place.

For the hybrid architecture, the answer lies in the appreciable memory and power savings from using DNNs versus LUTs. As such, this architecture may be largely beneficial for achieving objectives other than safety: indeed, at a higher level of abstraction, this pattern is effectively an instance of single channel design (Section 3.1.1). Thus, all the channels and the monitor/checking function inherit the totality of the assurance requirements for the function that is allocated to the pattern.

That, in turn, affects condition four above. Specifically, there is no relief in the level of assurance for the primary channel as in other architectural patterns involving a safe alternative function. In the hybrid architecture, assurance of the DNNs is expected to be shown using *learning assurance*, a new process being defined in (SAE and EUROCAE 2022) and first conceived in (EASA 2021b). The latter does not yet support high-criticality applications, while it remains to be seen whether regulators will endorse the former as an acceptable means of compliance to airworthiness regulations.

As such, for functions assigned higher DALs, the input partitioning and selection pattern will likely to be combined with other patterns such as RTA, backup parallel design, or *triple modular redundancy* (TMR) and the combination analysed together, to establish their suitability for use.

## 4 Concluding Remarks

Although others have assessed both hardware and software architectural patterns from a safety standpoint (Armoush et al. 2009, Hammett 2002), the integration of AI/ML was not a consideration. In (Armoush et al. 2009), the main safety focus was on achievement of reliability targets, related to so-called *safety integrity levels* (SILs), an orthogonal concept to development assurance levels (DALs) as we have considered in this paper.

The architecting problem when using ML/AI components has led to several variants of the RTA and system-level simplex architectural patterns—e.g., *certified control* (Jackson et al. 2021), and synergistic redundancy (Bansal 2022)—as well as proposals for novel architectures not conforming to previously established patterns or their combinations, e.g., *model-centred assurance* (Jha et al. 2020). Other architectural patterns established for functional safety have also been analysed from the standpoint of SIL allocation (Koopman 2021). However, the application domain for all of the above is autonomous road-vehicles.

Candidate architectures and patterns need to be assessed for their suitability in aviation applications so that guidance can be provided on how to justify credibility of the chosen form. This assessment will have to take into account compatibility with existing architectures, e.g., integrated modular avionics (IMA) used in aircraft systems. We have taken the first steps in this paper to identify and collate architectural patterns, both established and newly proposed, for including AI/ML in an aviation context, and assessed their potential suitability for use when viewed through the lens of safety assurance.

The DAL paradigm for sufficiency of safety assurance largely influences the choice of an architecture in an aviation system (in particular for aircraft systems), under the current certification regime. That is, DALs modulate how much assurance is required in proportion to the safety criticality of a function. They translate a level of assurance (e.g., Levels A - E) to the extent of development rigour necessary, which is codified in terms of process objectives. The higher the DAL, the greater the development rigour needed, and more evidence needs to be produced. Guidance does not currently exist for the nature of the evidence that will be needed to provide appropriate confidence that proposed architectures including ML/AI are fit for purpose. Although process assurance guidance for AI/ML-based products is being developed (SAE and EUROCAE 2022), the guidance it contains on architecture and architecting is largely implicit.[9]

The discussion earlier in this paper suggest that the prevailing architectural patterns (e.g., from ARP 4754) remain valid in principle when used with AI/ML-based complex functions. However, in practice they will need to be modified to address the nature of the failure modes of ML components. Such adjustments are also likely to be needed by other architectural patterns used in aviation (Hammett 2002) that we have not analysed in this paper. Although new architectural patterns have been proposed for integrating AI/ML, adopting them for use in aviation is far from straightforward. Indeed, in their current form, some may be unsuitable for use for high-criticality functions when confronted with the stringency of the associated assurance requirements. Moreover, the complexity required of components such as monitors may mean that the effort and cost of employing such architectures is increased.

---

[9] Two of the authors are members of the standardisation committee that is formulating the guidance in (SAE and EUROCAE 2022), and the associated technical exchanges have had a part in motivating us to craft this paper.

Our discussions lead us to the following recommendations:

1. Generation of a catalogue of architectural patterns and, potentially, combinations thereof for safety of systems containing untrusted technology—for systems in general, and aviation systems in particular.
2. An assessment framework by which the suitability of a proposed architectural pattern can be assessed for credibility during system development. For example, work is required to identify proportionate assurance requirements over the architectures and to determine whether additional guidewords are required for safety assessment when utilising AI/ML.
3. Development of assurance or confidence case patterns associated with the use of the architectural patterns in the catalogue.

Finally, this work is at an early stage. We welcome feedback on both the usefulness of these recommendations, and the ways to fulfil them.

**Acknowledgements**  Ganesh Pai contributed to this work under support from the System-wide Safety (SWS) project under the Airspace Operations and Safety Program of the NASA Aeronautics Research Mission Directorate (ARMD).

**Disclaimer**  The opinions, findings, recommendations, and conclusions expressed are those of the authors and do not represent the official views or policies of KBR, Inc., the National Aeronautics and Space Administration, and United States Government.